\documentstyle[aasms4,psfig]{article}

\newcommand{\kms}{{km s$^{-1}$}} 
\newcommand{\zabs}{{z$_{\rm abs}$}} 
\begin{document} 
 
\title{Chemical evolution of  Damped Ly$\alpha$ galaxies: 
The [S/Zn] abundance ratio at redshift $\geq$ 2
\footnote{Based on observations made with the William Herschel telescope 
operated on the 
island of La Palma by Isaac Newton Group in the 
Spanish Observatorio del Roque de Los Muchachos  of the 
Instituto de Astrof\'\i sica de Canarias, and on observations
collected with the ESO 3.6m telescope at the European Southern Observatory, 
Chile.}  
}

\author{Miriam Centuri\'on, Piercarlo Bonifacio, Paolo Molaro, and 
Giovanni Vladilo} 
\affil{Osservatorio Astronomico di Trieste, Via G.B. Tiepolo 11, 34131, Trieste} 
\authoremail{centurio@oat.ts.astro.it}

\begin{abstract} 
{
 Relative elemental abundances, and in particular the $\alpha$/Fe ratio, are
an important 
diagnostic tool of 
the chemical evolution of damped Ly $\alpha$ systems (DLAs).  
The S/Zn ratio  
is   not affected 
by differential dust depletion and  is an excellent estimator 
of the $\alpha$/Fe ratio. 
We report 6 new determinations of sulphur abundance in 
DLAs at \zabs $\geq$ 2 with already known zinc abundances.
The combination with extant data from the literature provides
a  measure of 
the S/Zn abundance ratio for a total of 11
high redshift DLA systems. 
The observed [S/Zn] ratios    
do not show the characteristic [$\alpha$/Fe] 
enhancement observed in   metal-poor stars of the Milky Way 
at comparable level of metallicity ([Zn/H] $\approx -1$). 
The behaviour of DLAs data is consistent with a general trend of 
decreasing [S/Zn] ratio 
with increasing metallicity [Zn/H]. This would be the first evidence of the
expected decrease of the $\alpha$/Fe ratio in the course of chemical evolution
of DLA systems.
However,
in contrast to what observed in our Galaxy, 
the $\alpha$/iron-peak ratio seems to attain solar values when the 
metallicity is still
low ([Zn/H]$\leq$--1) and  to decrease below solar values at higher metallicities. 
The behaviour  of the $\alpha$/Fe ratio
challenges the frequently adopted hypothesis that high redshift DLAs 
are progenitors of spiral galaxies and favours instead an origin in 
galaxies characterized by low star formation rates, 
in agreement with the results from imaging studies of low redshift DLAs, 
where the candidate DLA galaxies 
show a variety of 
morphological types including dwarfs and LSBs and only a minority of spirals.}  
\end{abstract}

\keywords{cosmology: observations --- galaxies: abundances --- galaxies: evolution
--- quasars: absorption lines } 
 
\section{Introduction}
 
Damped Lyman $\alpha$ systems  are  QSO absorbers  
with  the highest values of neutral 
hydrogen column density   (log\,$N$(HI) $\geq$ 20.3 atoms/cm$^{-3}$). 
The Ly$\alpha$ absorption profiles show  extended  
"radiation damping" wings (hence their name) and  are  
always associated with   narrow 
metal absorptions. 
Even though it is generally agreed that DLA  
absorbers at high redshifts originate in  proto-galaxies located in the direction of the
background QSO,  the  nature  
of such intervening galaxies is still subject of debate. 

The analysis of the kinematics
of the metal lines has   suggested  that DLAs   arise in 
massive rotating
disks  which are the  progenitors of the 
present-day spiral galaxies (Wolfe et al. 1995, Prochaska \& Wolfe 1997,1999).  
However, other works indicate that the observed kinematical properties can be
equally explained by low-mass proto-galactic objects  (Haehnelt, Steinmetz, \& Rauch 1998;
Ledoux et al. 1998). 

The imaging  of galaxies in the field of the background QSOs 
indicates that at \zabs $\leq$ 1 --- where this technique can be applied ---
the population of DLA galaxies is not 
dominated by a specific morphological type,  and, in particular, spirals 
constitute 
a small fraction of the sample (Le Brun et al. 1997; Rao \& Turnsheck 1998).  

Abundance studies of DLAs have also the potential  to  provide
independent clues to
understand the nature of  this class of QSO absorbers.   
Abundances determinations have   
already been obtained for about 60 systems, mainly at  z$_{\rm abs}>$ 1.7,
(Lu et al. 1996; Pettini et al. 1997,1999;
Prochaska \& Wolfe 1999, hereafter PW99). DLA galaxies show   
low metallicities --- typically 10\% of solar ---  comparable  
with the metallicity level measured in metal-poor stars of the Galactic halo.
Although the precision obtained in DLA's
abundance determinations  is remarkable and often comparable to that attained in the
Galactic ISM, the interpretation of the observed abundance ratios
has led to contradictory conclusions.

If DLA galaxies are protospirals and have experienced a chemical  evolution similar  
to that of our Galaxy, we expect to observe the elemental  
abundance pattern  typical of halo stars of  comparable metallicity. 
In particular,  we expect to observe the  enhancement
of $\alpha$ over iron-peak elements    ratio   
[$\alpha$/Fe]\footnote{Using the standard  
definition [X/Y]= log (X/Y) - log (X/Y)\sun}\,$\simeq$+0.5 characteristic
of Galactic metal poor stars  (Ryan, Norris, \& Beers 1996). 
The only [$\alpha$/Fe] ratio with a large number of
determinations   
in DLAs is [Si/Fe]. 
The    mean value  
$<$[Si/Fe]$>$ $\simeq$ +0.4$\pm$0.2  is consistent with the  
typical value of the Galactic halo stars and has been interpreted as evidence
for an origin of DLAs in proto-spirals (Lu et al. 1996, PW99). 
 However, also in the nearby interstellar medium one tipically 
finds $<$[Si/Fe]$>$ $\simeq$ +0.4 owing to differential elemental depletion  
onto dust grains (Savage \& Sembach 1996).
Vladilo (1998)  obtained  intrinsic solar ratios ([Si/Fe]$\approx$0)  
in DLAs after correction
for the differential elemental depletion. Also Pettini et al. (1999)
have recently found solar Si/Zn ratios in 3 DLAs by taking into account 
dust effects.
 
By using ratios of undepleted elements  such as [S/Zn],
Molaro et al. (1996)  
and Molaro, Centuri\'on,\& Vladilo (1998; hereafter MCV98) also 
obtained solar ratios, in spite of the low metallicity of the DLAs investigated.   
Since sulphur and zinc are essentially
undepleted in the interstellar medium,    the [S/Zn]   ratio
is probably the best diagnostic of the
[$\alpha$/Fe]  ratio available for DLAs. 
However, only a few measurements of   sulphur abundances  are available
owing to the difficulty of observing the SII resonant triplet 
in the Ly $\alpha$ forest. For this reason we have performed
a search for SII in DLAs. Here we present the results of
this search
in 7 DLAs
 with known
ZnII abundances. We obtain  
4 sulphur abundance measurements and 2 limits, which
allow us to enlarge significantly the sample of [S/Zn]  determinations.

\section{Observations and data reduction} 
 
The QSOs observed in the course of the present investigation  are
listed in Table 1, together with relevant information concerning 
the observations. The spectra of QSO 0013-004 and QSO 2231-0015 
were obtained  with the  CASPEC echelle spectrograph at the Cassegrain focus 
of ESO 3.6m  telescope at La Silla, Chile. The spectra of 
the remaining QSOs  
were obtained with the two arms ISIS 
spectrograph at the Cassegrain focus of the William Herschell Telescope  
(WHT, 4.2m.) at La Palma, Canary Islands.  
 
For the CASPEC observations we used the
echelle grating of 31.6 grooves/mm and a Tektronix CCD with 1024x1024  
square pixels of 24 $\mu$m in size.  The CCD was binned at a 
step of 2 pixels along the dispersion direction and the slit width was  
set at 2.1 arcsec 
in order to have the projection onto 2 binned pixels of 
the detector.  The slit width 
matched the seeing,   
which was around 2 arcsecs during the observations at La Silla.  
  
For the observations performed with
ISIS blue arm we used a 1200 grooves/mm grating coupled with an EEV 
CCD with  
2048 x 4200 square pixels of 13.5 $\mu$m in size.  
The CCD was binned in the spatial and dispersion 
directions at step of 2 x 2  pixels. The slit width was set at 
1 arcsec. This value still allows a correct sampling of the spectrum without  
any loss in resolution thanks to the small pixel size of the EEV CCD.    
  
The full width at half maximum of the instrumental profile sampled with 2
binned pixels, $\Delta 
\lambda_{\rm instr}$, was measured from the emission lines of the 
Thorium-Argon lamp (CASPEC)  and from the Copper-Argon plus  
Copper-Neon lamps (ISIS blue arm) recorded contiguously to each 
target exposure.  The resulting resolving power R= $\lambda$/$\Delta 
\lambda_{instr}$ was R $\simeq$ 19000 for CASPEC spectra  
and $\simeq$ 5000 in the sulphur region of ISIS blue data, corresponding to  
a velocity 
resolution of $\Delta$v $\simeq$ 16 \kms and  60 \kms,
respectively. 
 
The data reduction was performed using the ECHELLE (CASPEC spectra) and  
LONG (ISIS spectra) routines 
implemented  in the software package MIDAS  developed at ESO.     
The first steps of the   reduction procedure ---  including
flat-fielding, cosmic ray removal, sky subtraction,  optimal extraction, and
wavelength calibration  ---
were performed separately on the different spectra of each QSO. 
Typical internal errors in the wavelength calibrations are of 
$\simeq$ 4\,m\AA\, for CASPEC spectra and $\simeq$30\,m\AA\, for ISIS blue data. 
  
The observed wavelength scale of the spectra was then
transformed into vacuum, 
heliocentric wavelength scale.  At this point the different spectra of 
each QSO were  
averaged, using as  weights the   continuum levels of the exposures. 
Finally, for each spectral range under study the local continuum was determined
in the average spectrum by using a spline to smoothly connect the  
regions free from absorption features.  The final spectrum used for the analysis
was obtained by normalising the average spectrum to these continua. 
The signal-to-noise ratios per pixel of the extracted final spectra,
estimated from the $rms$ scatter of the continuum near the absorptions 
under study, are typically in the range between 10 and 25.

In addition to the S\,II triplet at 1254\AA,
our data also cover, in general, the spectral regions of NI multiplets (1134, 
1200 \AA), and the transitions of OI(1302, 1355\AA), SiII (1190, 1193,  
1260, 1304, 1526\AA) and  
FeII (1121,1125,1127,1133,1143,1144 \AA).
In this paper we focus our attention on the new measurements of the SII triplet.
When possible, we used other lines to obtain information  
that could be used to constrain the SII column densities. 
We do not report OI column densities because the transition at 
$\lambda$ 1302\AA\ is  heavily saturated, while 
the extremely weak OI 1356 \AA\  line is not detected and provides 
no stringent upper limits.  
Nitrogen abundance determinations 
in these DLAs will be  discussed in a subsequent paper.

\section{Column densities} 

Column densities have been obtained by
fitting theoretical Voigt profiles 
to the observed absorption lines   
via $\chi^2$ minimization. 
This step was performed   using the  
routines FITLYMAN (Fontana \& Ballester 1995) included in the MIDAS 
package. During the fitting procedure the theoretical profiles were convolved    
with the instrumental point spread function 
modeled from the analysis of the emission lines of the arcs.  
Portions of the profiles contaminated 
by intervening Ly $\alpha$ absorbers 
were excluded from the fit.  

The FITLYMAN routines determine
the  redshift, the column density (atoms cm$^{-2}$), and the
broadening parameter ($b$-value)\footnote
{The broadening parameter is defined as $b = 2^{1/2} \sigma_v$,
where $\sigma_v$ is the gaussian velocity dispersion for Doppler
broadening. 
}
 of the 
absorption components, as well as the fit errors for 
each one of these quantities. In addition, we estimated errors due to the
uncertainty in the continuum placement, as we explain in the rest
of this section. 
 
Measurements of SII column densities  were derived 
from the three lines at $\lambda$$\lambda$1250.584, 1253.811, 1259.519 \AA\,    
with oscillator strengths f$_{\rm {\lambda}}$=0.00545, 0.01088, 0.01624 
respectively. 
All the atomic data used in this  work are from Morton (1991). 
The SII triplet  
is generally  unsaturated, but can be   
contaminated by the Lyman $\alpha$ forest and we have been able 
to measure the sulphur column density only in four out of the seven  
DLAs under study.

In Table 2 we give our derived  
SII column densities and $b$-values.  
No sulphur  abundance have 
been previously reported for the DLA systems under study. 
We now comment briefly the column density measurements of each DLA, starting
with the two systems observed at higher spectral resolution (CASPEC data). 
For the ISIS data we first discuss those for which a determination of  
SII column density has been possible.

\subsection{System at \zabs=1.9730 toward QSO 0013-004} 
 
For this absorber
we derived the SII column density from the fit to the  
$\lambda$1253\,\AA\ transition. The bluest transition at $\lambda$1250\,\AA\ ---  the weakest one 
of the triplet ---  is quite noisy  
since  it is located in the first order of the echellogram where the S/N 
ratio is low. 
The reddest  
transition at $\lambda$1259\,\AA\  is completely blended, as one can see  in Fig. 1a. 
We remark that the SII 1253\,\AA\  line is found exactly at the same redshift
(\zabs=1.9730) as the ZnII and CrII absorptions studied by Pettini et al. (1994).   
The $\lambda$1250\AA\  transition, in spite of the low S/N ratio, is very 
well matched by  
the synthetic spectrum built with this redshift value (solid line in Fig. 1a).

Our best fit to the SII 1253\,\AA\  absorption gives log N(SII)=14.86 and 
b=24.5 \kms.
The rest-frame equivalent width of the same line, EW$_{\rm rest}$= 81 m\AA, 
yields log N(SII)=14.74 by using the linear part  
of the curve of growth (COG) which is indicative within the errors of a 
linear regime. In fact we obtained an acceptable fit to the SII 1253 
absorption for a wide range of b-values, beeeing 
b=14 and b=35 the extreme  
values and
in both cases the resulting SII column density is
consistent with the best fit value within the errors.
We explored the  
uncertainty in the column density coming 
from the continuum tracing by shifting the continuum $\pm$1$rms$ 
and repeating the fit in each case.  
We obtained the best fits for log N(SII)=14.77 (b=20.4), and  
log N(SII)=14.91 (b=24.6) 
for the lower and upper continuum 
respectively.  We adopt log N(SII)=14.86 ($\pm$0.12) and b=24.5($\pm$10) 
given in Table 2, which takes into account the largest excursion of the 
SII column 
density values. 
 
Unfortunately the SiII transitions available in our spectrum    
($\lambda$$\lambda$1260, 1304, and 1526 \AA) are saturated or blended, making 
impossible a determination of the SiII column density for this system.

\subsection{System at \zabs=2.0662 toward QSO 2231-0015} 
 
Only SII 1253\,\AA\  absorption is available  in this DLA, since 
the other two transitions of the triplet are contaminated by the   
Lyman $\alpha$ forest (see upper pannel of Fig. 1). 

The fitted SII 1253\,\AA\  line is observed at the same redshift
(\zabs=2.0662) as found by Lu et al. (1996) and PW99 
in their study of this system. Moreover part of our wavelength range is 
also covered in the spectrum of PW99, and features in both spectra
like OI 1302 and SII 1304  
occur also at the same velocity position.

The SII absorption presents a high degree 
of saturation and only with an  
independent 
estimation of the $b$-value it would be possible to attain  a reliable    
estimate of the SII column density.
Unfortunately, no information is available 
on $b$ value for this DLA, 
since abundances before this work  
have been obtained by using the  
opacity method. 
The NiII (1370\AA) and CII (1335\AA) absorptions observed in the  
higher S/N spectrum of PW99  
are not detected in our spectrum making it impossible to constrain 
the $b$ parameter 
in this system.  
We estimated log N(SII) $>$ 14.90, a  
lower limit obtained by using the rest-frame equivalent width and the linear part 
of the COG.

\subsection{System at \zabs=2.3745  toward QSO 0841+129} 
 
This is the brightest target observed. The spectrum of this BL Lac object  
discovered by C. Hazard shows  
two DLAs previously analyzed by Pettini et al. (1997) and shown in  Fig. 2a. 
Our best fit to the lower redshift damped absorption at \zabs=2.374 gives  
log N(HI) = 20.96$\pm$0.10 in perfect agreement with log N(HI)=20.95$\pm$0.10  
reported by Pettini et al. (1997). 
 
For this \zabs=2.3745 system     
a feature at 4 $\sigma$ significance level is observed at the expected  
redshifted  
position of  
the  strongest triplet transition SII 1259\,\AA\  (vertical arrow in 
Fig. 2b). The blue ISIS spectrum of Pettini et al. (1997; see their Fig. 1)    
recorded at about 
half of our resolution also shows this feature  
indicating that it is real.  
 
The SII 1259\,\AA\  transition is observed in the red 
wing  of the higher redshift  
damped Ly $\alpha$ absorption at \zabs=2.476,  
which unfortunately precludes the detection of the other two bluer  
absorptions of the SII triplet. 
Our best fit to this damped absorption (\zabs=2.476) shown with solid  
and dotted  
lines 
in Fig. 2a,b gives log N(HI)= 20.83$\pm$0.10 
in good agreement with log N(HI)=20.79$\pm$0.10 reported by  
Pettini et al. (1997). 
We re-normalized this portion of the spectrum to the Ly $\alpha$  
profile  before fitting the SII line.  
For this system there is not independent information on the $b$-value  
from the literature. In order to constrain the 
SII column density  we estimated the $b$- value in this line of sight
from the analysis of the  
FeII $\lambda$$\lambda$ 1121, 1125, 1127,1133,1143,1144 \AA\ 
transitions. We obtained log N(FeII)=14.83$\pm$0.15 and b=20$\pm$3 \kms  
(see Fig. 2c). 
The errors in the FeII column density and $b$-value  
take into account the fit errors and in the error due to the uncertainty of the
continuum tracement.  
 
By fixing $b$(SII)=$b$(FeII)= 20 \kms\ and \zabs=2.3745 ---  
at which we observe 6 FeII absorptions --- in the SII fitting procedure
we obtained log N(SII) = 14.92$\pm$0.09. In Fig. 3a this fit is shown  
with a solid line overimposed to the spectrum re-normalized to the 
Ly$\alpha$ absorption. 
Changing the $b$-value by $\pm$3 
\kms\ affects the SII column density  only  by $\leq$ 0.02 dex. 
Moreover if we use b=13 \kms obtained from the analysis of 
the NI 1134 \AA\ and 1200 \AA\  multiplets  
we obtain log N(SII) = 14.89$\pm$0.14 still consistent with 
the column  
density obtained for $b$=20 \kms. These results clearly indicate that the 
SII absorption under study is unsaturated and this is reinforced by the fact
that by using  
the rest equivalent width (EW$_{\rm rest}$ = 175 m\AA)  
over the linear part of the COG 
we obtained again log N(SII) = 14.89.  
We estimated the error  due to    
the continuum placement by fitting the SII absorption in the  
spectra normalized to the  
local continua shown with dotted lines in Fig. 2a, and we obtained 
$\epsilon$$_{\rm {log N(SII)}}$= $^{+0.16}_{-0.21}$. This error dominates
the error budget and we adopted  
log N(SII) = 14.92 $^{+0.16}_{-0.21}$. 
We remark that in case we consider the SII absorption non detected, we can use
log N(SII) $<$ 14.92 as a conservative 4$\sigma$ 
upper limit.
Even in this case 
the result would not affect the main conclusion of the
present work, as we discuss in Section 4. 
Nevertheless as it has been explained above there are at least
three good reasons which make this detection reliable: i)
the feature seems to be also present in the spectrum of Pettini et al. 1997,
ii) the feature is observed at \zabs=2.3745 the same redshift at
which we observed six single-component transitions of FeII shown in Fig. 2c,
iii) the single component-structure and the redshift are confirmed  
by the absorptions of SII 1808, CrII 2056,2062,2066, ZnII2062
observed in the higher resolution and S/N HIRES-Keck spectrum of PW99.

\subsection{System at \zabs=2.4762  toward QSO 0841+129} 
 
As can be seen in Fig. 2a,b the SII triplet 
of this system is overimposed to a smooth, broad absorption which is also seen in the spectrum 
of Pettini et al. (1997).  
We   investigated the possibility that this broad absorption may be due to 
Ly$\alpha$ line-locking   
with the velocity separation of CIV or SiIV resonance doublets 
since this process is associated  
with absorption systems at \zabs $\simeq$ z$_{\rm {em}}$ of the QSO (Srianand, 1999). 
This is the case here since z$_{\rm {em}}$ $\simeq$ 2.5   
have been estimated from the     
onset of Ly $\alpha$ forest (Pettini et al. 1997).  However,
from the line-locking process  one would expect a velocity separation 
between the  
broad profile and the   
\zabs=2.4762  Ly$\alpha$ absorptions correspondent to the  
velocity separation between the two lines of the CIV  or SiIV   doublet   
of that system. 
This is not the case here, at least for the CIV or SiIV doublets. 
In any case,   the presence of the 
broad feature does not affect  our   measurements  because the SII 
absorptions are
thin, as expected for DLA metal lines, and are clearly seen distinguishable 
from
the broad feature. 
In order to analyze the SII triplet we  renormalized this portion of the 
spectrum to the broad  profile   shown in Fig. 2b. 
The SII 1253\,\AA\  transition is heavily blended  and we excluded this feature 
from the fitting procedure.    
Our best fit to the SII 1250\,\AA\  and 1259\,\AA\  transitions gives  
log N(SII) = 14.81,  b=13.5 \kms and  \zabs= 2.4762. 
In Fig. 3b we show 
the synthetic  
spectrum of the SII triplet computed with these parameters.
We remark that redshift obtained from the SII triplet is in perfect agreement
with the one found by PW99 for different single-component metal 
absorptions observed in their 
Keck 
spectrum, and hence enhancing the realiability of the SII detections.
Pettini et al. 1999 gave \zabs= 2.4764 for this system but it was obtained 
just from the fit to the wide damped Lyman $\alpha$ absorption.

By using the rest equivalent width (EW$_{\rm rest}$ = 48 m\AA) of the weakest  
1250 \AA\ transition 
over the linear part of the COG 
we obtained esentially the same column density 
log N(SII) = 14.80 indicating that this transition is not saturated.  
Again the major uncertainty in the column density comes from the  
continuum placement. We renormalized the spectra to the  
upper and lower continua shown with dotted lines in 
Fig. 2b and we measured for SII 1250\,\AA\  absorption  
EW$_{\rm rest}$= 70 and 30 m\AA\ which yield log N(SII)=14.96 and 14.60 
atoms/cm$^{-2}$ 
respectively. 
We adopted log N(SII) = 14.81$^{+0.15}_{-0.21}$ given in Table 2.

\subsection{System at \zabs= 1.999  toward QSO 1215+333}

Also in this system the SII triplet is  observed overimposed on a wide  
absorption (see Fig. 2d). In this case
an origin in the  line-locking  process   is unlikely because there is a  
signicative difference between   \zabs\   and  z$_{\rm {em}}$    
(see Table 1).  
To analyse the SII triplet we have renormalized this portion of the spectrum 
to the broad absorption (Fig. 3c). 
The SII 1253\,\AA\  transition though partially blended is the only 
feature that can be used to derive the column density.  
 
The SII 1259\,\AA\  transition  
is heavily blended and our best fit to the SII 1253\,\AA\  line gives 
log N(SII)= 15.11 for b=55 \kms. A synthetic SII spectrum built with these  values  
is shown with a solid line overimposed to the observed spectrum in Fig. 3c. 
In order to explore the degree of saturation of this absorption we  
performed the fit for a  large range of $b$ values.  
We obtained $b$=70 \kms (log N(SII)=14.99) and $b$=18 \kms (log N(SII)=15.23)  
the maximum  and minimum  
b values which can give an acceptable fit to the observed profile. 
Again, the column density does not depend very much on the $b$-value, 
indicating a low degree of saturation. In fact the SII 1253   
EW$_{\rm rest}$=0.150 \AA\ gives logN(SII)=14.99
over the linear part of the  COG.  
Also in this case the major source of uncertainty in the  column density  
comes from the continuum placement. In order to estimate this error we
re-normalized the spectrum to the  continuum positions  shown  
with dotted 
lines in Fig. 2d. With the upper continuum 
we obtained a minimum $b$=15\kms\  
for which an acceptable fit can be obtained, yielding logN(SII) = 15.36. 
In this case the 1253 line presents a certain degree of saturation since 
the rest equivalent width over the linear part of the COG gives  
logN(SII) = 15.10. 
For the lower continuum the column density does not depend on the $b$-value 
and we obtained log N(SII)=15.00$\pm$0.02 for 18$\leq$b$\leq$50 \kms. 
We  obtain therefore  for this system  
log N(SII) =  15.11$^{+0.25}_{-0.12}$. 
Nevertheless, since this determination relies on just one absorption
which is partially blended and no other single metal absorptions are 
available in our spectrum to confirm the redshift, we adopt the most 
conservative result by considering  
the sulphur abundance as an upper limit log N(SII) $<$ 15.11+0.25, and we 
remark
that this does not change the main conclusion of this work,
as discussed in Section 4.

\subsection{System at \zabs=2.1408 toward QSO 0149+335} 
 
None of the three absorptions of the SII triplet are detected.  
Their expected redshifted positions are marked with dashed vertical lines 
in Fig.3d. The  
low signal-to-noise ratio in this spectral range (S/N$\simeq$6) yields 
a poor stringent upper limit on the SII column density and hence on  
the abundance of 
this element (see Table 2 and Table 3.)

\subsection{System at \zabs= 2.4658 toward QSO 1223+178 }

In Fig. 3e we show the SII portion of the spectrum already normalized to the 
broad absorption. The vertical dashed lines show the expected positions 
of the triplet.    
The SII 1253\,\AA\  line if 
present is contaminated by a cosmic ray, and the SII 1259\,\AA, if present, 
is heavily blended with a Ly$\alpha$ interloper. 
The weakest 1250\,\AA\  absorption seems undetected in our S/N $\simeq$ 7 
spectrum. 
If this is the case we obtain log N(SII)$<$15.18 and an  
upper limit [S/Zn]$\leq$--0.01 by using the ZnII column density given 
by Pettini et al. (1994). However  the poor quality of the spectrum in the 
SII region and the difficulty in positioning the continuum precludes 
a reliable analysis of the SII  
triplet in this system and for that reason we 
have omitted this DLA  from the rest of the present study.

\section{Discussion}

\subsection{ Abundances of $\alpha$ and iron-peak elements  } 

In Table 3 we list the  abundances of the iron-peak elements 
Fe, Zn and Cr as well as  the abundances of the $\alpha$ elements S and Si 
for the DLA systems under investigation. 
 
Iron-peak elements are produced in nuclear statistical equilibrium and 
are expected to trace each other in the course 
of  chemical evolution.  
Observations of metal-poor stars in the Galaxy  
confirm that Cr and Zn follow closely  Fe 
in essentially solar proportions down to very low metallicities, 
although Cr deviates from this behaviour becoming slightly underabundant 
compared to iron   
at about [Fe/H] $\leq$ -2 (Ryan et al. 1996;
Sneden, Gratton \& Crocker 1991)
   
DLA systems,  which have metallicities [Zn/H] $>$--2,
show instead  systematic differences among the iron-peak elements,
with  Zn more abundant than Cr and Fe. Abundances of these elements in the
DLAs under study are also given in Table 3.  
The systematic difference between Zn and Cr is  attributed to differential 
depletion of these two elements from the gas phase to dust grains 
(Pettini et al. 1994, 1997). In fact, enhanced [Zn/Cr] and [Zn/Fe] ratios 
are observed 
also in the nearby interstellar medium, which is expected to have solar 
chemical composition, and are attributed to differential dust depletion,
being Zn almost undepleted
(Roth \& Blades 1995, Savage \& Sembach 1996). 

The presence of  dust depletion may also affect the analysis of some 
$\alpha$-elements such as silicon.
Silicon is the $\alpha$  element with the  largest  
number of measurements in  DLA systems   
(Lu et al. 1996, PW99), however can be depleted up to  1 order of magnitude 
in the Galactic interstellar medium (Savage \& Sembach 1996). 
Sulphur also shows in Galactic metal-poor stars the typical enhancement  of  
$\alpha$-elements with 
[S/Fe] $\simeq$ +0.4/+0.6  (Francois 1988), 
but  contrary to Si,  sulphur 
is undepleted from gas to dust.

The fact that typical 
depletions of S and Zn in Galactic interstellar clouds  are in the ranges 
[-0.05,0.0] 
and [-0.25, -0.13] dex respectively  
(Savage \& Sembach 1996;  Roth \& Blades 1995), makes the 
[S/Zn] ratio  a reliable dust-free diagnostic tool of  
the [$\alpha$/iron-peak] abundance 
ratio in DLAs.

In Table 4 we compile all the extant [S/Zn]  measurements in DLA systems. 
For the sake of comparison we also give, when available, the [Si/Fe] ratios 
for these DLAs. One can see that    
while the [Si/Fe] ratios emulate 
the typical halo-like abundance pattern ([$\alpha$/Fe] $\simeq$ +0.5), 
the [S/Zn] ratios do not show evidence for $\alpha$/Fe enhancement.   
Following  Vladilo (1998) we have corrected, when possible
\footnote{For DLAs with available abundances of both Fe and Zn},
the observed [Si/Fe] ratios
from dust effects. These values are  
also reported in Table 4, and show   that, when dust correction is 
quantitatively taken 
into account, the corrected ratios, [Si/Fe]$_{corr}$  have  
lower values similar  to the [S/Zn] ratios. 
This result  confirms  that dust  plays an important role 
in the observed 
[Si/Fe] overabundance and suggests that the assumptions adopted in 
the dust correction method seem to be appropiate for DLAs.

The \zabs=1.973 DLA system toward Q0013-004 included in our sample
is one of the few DLAs for which molecular hydrogen  
has been detected (Ge \& Betchold 1997). 
The presence of molecular gas is   
indicative of an environment hospitable to grains and we should expect  
a large depletion of refractory elements. This  is indeed  confirmed
by the 
much higher abundance of Zn compared to Fe (Table 3).
In a dust-rich DLA      
Zn can be expected to be somewhat depleted --- in a higher proportion than S  
as it occurs in the galactic ISM --- and  a  slight enhancement of the 
[S/Zn] ratio could be expected. 
In order to quantify this effect  
we corrected the ratio [S/Zn]$_{\rm obs}$=--0.39 measured in this system 
following  
Vladilo (1998) and we obtained [S/Zn]$_{\rm corr}$ = --0.41. 
The difference is very small and lower than  the typical measurement errors, 
supporting the assumption 
that the S/Zn ratio is a reliable indicator
of the $\alpha$/iron-peak ratios in DLAs also in the presence of a
significant amount of  dust.

Besides the problem of differential depletion,  the 
ionization balance could also
affect  the measurement of the elemental abundance ratios as discussed
recently by Howk \& Savage (1999). 
From the presence of Al III at the same radial velocity of low 
ions in DLA systems,  these authors argue that ionization effects can also produce an 
apparent enhancement of  the [Si/Fe] ratios measured  just from 
Si II and Fe II lines (i.e. from only one ionization state).
Dust or ionization effects are both threfore
going in the direction of producing an enhancement of the Si/Fe. 
However, from an analysis of literature data we  do not find evidence for 
a relative [Si/Fe] enhancement at the highest 
values 
of the ionization ratio Al III/Al II in the 5  DLA systems  
with available measurements of Al II, Al III, Si II and Fe II column densities.
 
In a separate work we show that the presence of ionized gas surrounding 
the HI regions in DLA systems should affect only marginally  
the relative abundances measured from low ions 
(Vladilo et al., in preparation).

\subsection{Implications for solar $\alpha$/Fe ratios}

In the early stages of the chemical evolution of galaxies the abundances 
are dominated by  
Type II SNae 
products, richer in $\alpha$-elements
yielding an enhancement of the
$\alpha$ elements  over iron-peak elements. 
At later stages  
of evolution 
the contribution of Type Ia SNae, richer in iron-peak elements,  
reduces the [$\alpha$/Fe-peak] ratio.  
The precise timing in which the products of
Type Ia  SNe become important depends on the star formation rate
and on the initial mass function. 
The [$\alpha$/Fe] ratio is therefore a primary indicator
of the type of chemical evolution and can be used to understand the
nature of DLA galaxies. 
 
In Fig. 4 we    
show the [S/Zn] measurements  in DLA galaxies.
 Our sulphur abundances are represented by squares which also include  
the  
DLAs at \zabs = 2.309  towards QSO 0100+1300 (PHL 957) and \zabs=3.025 towards 
QSO 0347--3819 discussed in  
MCV98. Triangles represent sulphur abundances 
from literature (see Table 4 for references). 
For the sake of comparison we also show in the figure the [S/Fe] ratios  
measured in Galactic metal-poor stars by
Francois (1987,1988), with star symbols,
and by 
Clegg et al. (1981), with  asterisks. 

Even at first glance it is clear that the [S/Zn] 
ratios in DLAs  
do not show the $\alpha$-enhancement 
seen in Galactic metal-poor stars and that are located in a different sector of 
the diagramme. This result, already  advanced in MCV98,
is now rather firm  since it is based on  6 new measurements 
of DLA systems which sample a wide 
range of metallicities. 
The 4 limits but one shown in Fig. 4, are also consistent with this result. 
The exception is the system at \zabs=2.476 towards Q0841+129 giving
[S/Zn]$>$0.2, which is therefore
consistent with an intrinsic $\alpha$-enhancement.
The above results are robust against a possible 
contamination of the sulphur absorptions by the 
Lyman $\alpha$ forest.
Should the Lyman $\alpha$ forest contaminate some of the detected SII features, 
the real [S/Zn] would be even lower and this would reinforce the result.  

The [S/Zn] ratios in DLA systems are significantly 
lower than the Galactic ones at comparable metallicities 
and suggest 
that DLA galaxies have undergone a different chemical evolution from that of  
the Milky-Way.
The frequently adopted hypothesis that DLAs are progenitors of the spiral 
galaxies, 
apparently supported by the observed [Si/Fe] ratios, is not confirmed  
when  a dust-free diagnostic as [S/Zn] is considered. Results based on [Si/Fe]
measurements should be treated with caution, unless dust depletion is
properly taken into account.

Some of the [S/Zn] ratios, far from showing an enhancement   respect 
to the solar value, show instead negative values. These cases happen at the 
highest values 
of metallicity in our sample, while the highest [S/Zn] values 
are observed at the lowest metallicities. 
In particular the lower limit suggestive of intrinsic enhancement
(\zabs=2.476 in Q0841+129) is found at the lowest value of [Zn/H] in 
our sample.
Therefore,   
the [S/Zn] abundances in DLAs are consistent with a general trend of 
decreasing ratio with increasing metallicity.
The limited amount of data are insufficient to
firmly establish the presence of a correlation, 
which only can be considered with the present data if
we use the [S/Zn] values in DLAs at \zabs = 2.374 towards QSO 0841+129
as a measurement and not as an upper limit.
In that case a linear regression to the 
data yields a correlation coefficient of r=--0.77, obtained only with 5 data 
points (see Fig. 4). If confirmed by a larger data sample the trend would be 
the first observational evidence of the expected decrease of 
the $\alpha$/Fe ratio in DLAs during the course of chemical evolution.
At variance with what observed 
in our Galaxy, however,  
the $\alpha$/iron-peak ratio attains solar values 
at low metallicity ([Fe/H] $\approx -1$) and decrease further at higher 
metallicities.  This trend  is the one predicted 
by chemical evolution models of  
galaxies with a low star formation rate (Matteucci et al. 1997)
and, if confirmed by future observations, it  would have an 
important implication on the origin of DLA systems.   
Chemical evolutionary models predict [$\alpha$/Fe]$\simeq$ 0 at low  
metallicities, ([Fe/H] $\leq$ -1) when star 
formation proceeds in bursts separated by quiescent periods, as happens 
in dwarf galaxies, and when star formation  
is not as fast as in our Galaxy, as it happens in LSB galaxies and in the 
outer regions of disks (Jim\'enez et al. 1999). In these galaxies
the metal enrichment is so slow that 
Type Ia supernovae have enough time to evolve and enrich the medium 
with iron-peak elements,    
in such a way as to  balance the  $\alpha$-elements  
previously produced by Type II supernovae, when the overall  
metallicity is still low.
In any case by considering the sulphur abundance in 
DLAs at \zabs = 2.374 towards QSO 0841+129
an upper limit we may still 
conclude that the [S/Zn] ratios in DLAs are markedly different from 
those observed in our Galaxy at comparable metallicities, implying
a different chemical history.

In Fig. 5 the  [S/Zn]  are plotted versus the  absorber redshift z$_{\rm abs}$. 
Chemical evolution effects should in general decrease the [S/Zn] 
with cosmic time and
one would expect a positive trend with  z$_{\rm abs}$. 
Our sample, however, 
does not show such a correlation.  
Pettini et al. (1997, 1999) do not find either any trend of the 
[Zn/H] ratio with z$_{\rm abs}$, contrary to the 
expectation of a general increase of metallicity with cosmic time.
A significant spread of [Zn/H] abundances at a given redshift
is expected  when different formation redshifts or
spatial gradients  are considered in modeling the
intervening galaxies (Jim\'enez et al. 1998). 
Thus, different epoch of formation and different regions within
a galaxy could be responsible also for the lack of correlation 
between the [S/Zn] ratios and redshift.
The fact that we possibly detect a trend with [Zn/H], but not with \zabs, 
suggests that metallicity is a better indicator of evolution since, 
contrary to redshift, it is independent of the epoch of formation
of the individual galaxies. We do not exclude however that evolution with 
redshift can be detected when the data will have a better redshift coverage.

It is worth mentioning that also in the Milky Way there are some
measurements of [$\alpha$/Fe-peak] ratios not enhanced at
low metallicity. These cases are found among halo dwarfs,
but are extremely rare. 
Carney et al. (1997) found  [Mg/Fe] = --0.31  in the star BD +80 245,  
while Nissen \& Schuster (1997) found  [Mg/Fe]  
ratios ranging  from -0.1 to 0.2 in their stars.
These stars are all characterized  by large apogalactic distances and
the unusual abundance ratios have been interpreted as the chemical signature of
a merger or accretion events. The stars before the merging or accretion
process  are thought to belong
to a satellite galaxy which  experienced a different chemical evolution
history than the Milky Way. The presence of these cases do not imply
therefore a connection between what we observe in DLA galaxies and the typical
behaviour of Milky Way chemical evolution.

We remark that also the  nitrogen abundances in
DLAs do not seem to follow the behaviour of Galactic metal-poor stars
when dust free ([N/S]), or dust-corrected ([N/Fe$_{corr}$]) ratios
are used to determine the nitrogen relative abundances 
(Lu, Sargent, \& Barlow 1998;  
Centuri\'on et al. 1998).

We conclude that the DLA galaxies do not show the abundance properties
usually expected for the progenitors, of  
a spiral galaxy as the Milky Way. The unusual abundance ratios
suggest that 
the DLA galaxies  are  objects with low, or episodic, star 
formation rates
such as LSB or dwarf galaxies.  Part of the DLAs may be
proto-spirals, for which the line of sight samples
the outer regions, which are known to have a slower
evolution than the internal regions of the disks.
 
These indications on the nature of DLA systems based on chemical abundances  
are in agreement with the results based on 
imaging studies 
at low redshifts, where the candidates DLA galaxies show a variety of  
morphological types including dwarfs and LSBs, while spirals are not the  
dominant contributors (Le Brun et al. 1997, Rao \& Turnshek 1998). 
Nevertheless, it is important to remark that the
spectroscopic  sample of DLA systems  
is probably biased  
against detection of spirals, since high column density clouds located in 
environments with relatively high metallicity and dust  
can be missed owing to   
obscuration of the background QSO (Pei et al. 1991, Vladilo 1999).

\clearpage 
 
\figcaption{SII triplet of z$_{abs}$=1.9731  and  
z$_{abs}$=2.0662 DLA systems observed in the CASPEC normalized spectra 
of QSO 0013-004 and QSO 2231-0015, respectively.  
Smooth lines: synthetic spectra obtained from the fit of the $\lambda$1253 \AA\  
transition of the SII triplet (see text for details). 
\label{fig1}}

\figcaption{Normalized portions of the ISIS blue spectra of
QSO 0841+129 and QSO 1215+333. (a) The two DLA systems 
towards QSO 0841+129. Solid and dotted lines show the synthetic damping
profiles corresponding to the column density of neutral hydrogen
given in Table 1. (b) SII 1259 \AA\ transition of \zabs=2.3745 DLA system,  
and SII triplet of DLA at \zabs=2.4762 towards QSO 0841+129.
The left single arrow  indicates the position of SII 1259 at \zabs=2.3745 
obtained 
from the six FeII absorptions and in perfect agreement with the value
given by PW99. The three arrows in the 
right side are plotted at the redshifted
position of the SII triplet for the DLA system at \zabs=2.4762. 
Solid and dotted lines show
the damped profile and the continua to which this portion of the spectra 
have been re-normalized in order to measure the SII column densities in 
both DLA systems. (c) FeII lines of \zabs=2.3745 DLA system in the 
normalized spectrum of QSO 0841+129. Solid lines show the synthetic 
spectrum obtained from the fit of all the FeII absorptions contemporaneously.
(d) Normalized portion of QSO 1215+333 spectrum in the region of the
DLA absorption and SII triplet. The solid and dotted lines are 
the continua to which the spectrum have been re-normalized in order
to analyzed the SII absorptions.
\label{fig2}}

\figcaption{Normalized blue ISIS spectra  
of the QSOs in our
sample centered in the SII triplet region. 
The solid lines are the synthetic SII spectra resulting from the best fits
to the observed profiles. Dashed vertical lines in panels d) and e)
indicate the expected positions of the undetected SII triplet.   
\label{fig3}}

\figcaption{ 
Measurements of [S/Zn] ratio in DLA systems. {\bf Squares}: Sulphur abundance
determinations obtained by our group including 
\zabs=2.309 towards QSO 0100+1300 and \zabs=3.025 towards 
QSO 0347--3819 given in MCV98.  	
Zinc abundances are from the literature (see references in Table 2).
{\bf Triangles}: Sulphur and Zinc abundances
determinations from the literature (see references in Table 4).
{\bf Stars and asterisks}: [S/Fe] ratio measured in Galactic stars by
Francoise 1987, 1988 and Clegg et al. 1981 respectively.
{\bf Dashed line}: Result of a linear regression of [S/Zn] versus [Zn/H] 
measurements in DLA systems including as measurements zabs=2.374 
towards QSO0841+129. Limits are not included 
in the correlation analysis.  
\label{fig4}}

\figcaption{ 
The [S/Zn] abundance ratios in DLAs plotted versus matallicity [Zn/H] 
\label{fig5}}

\psfig{figure=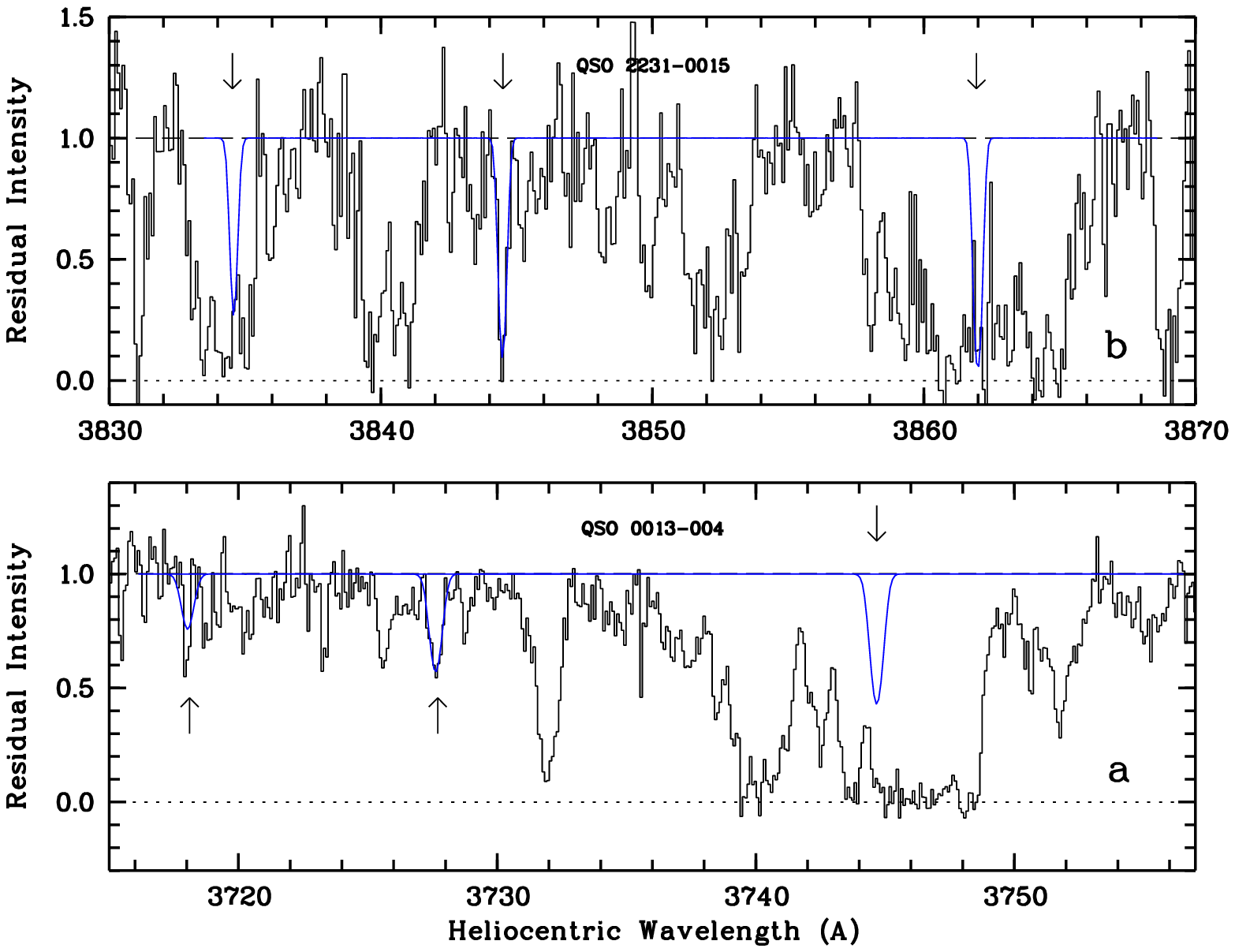,height=\vsize}  
\psfig{figure=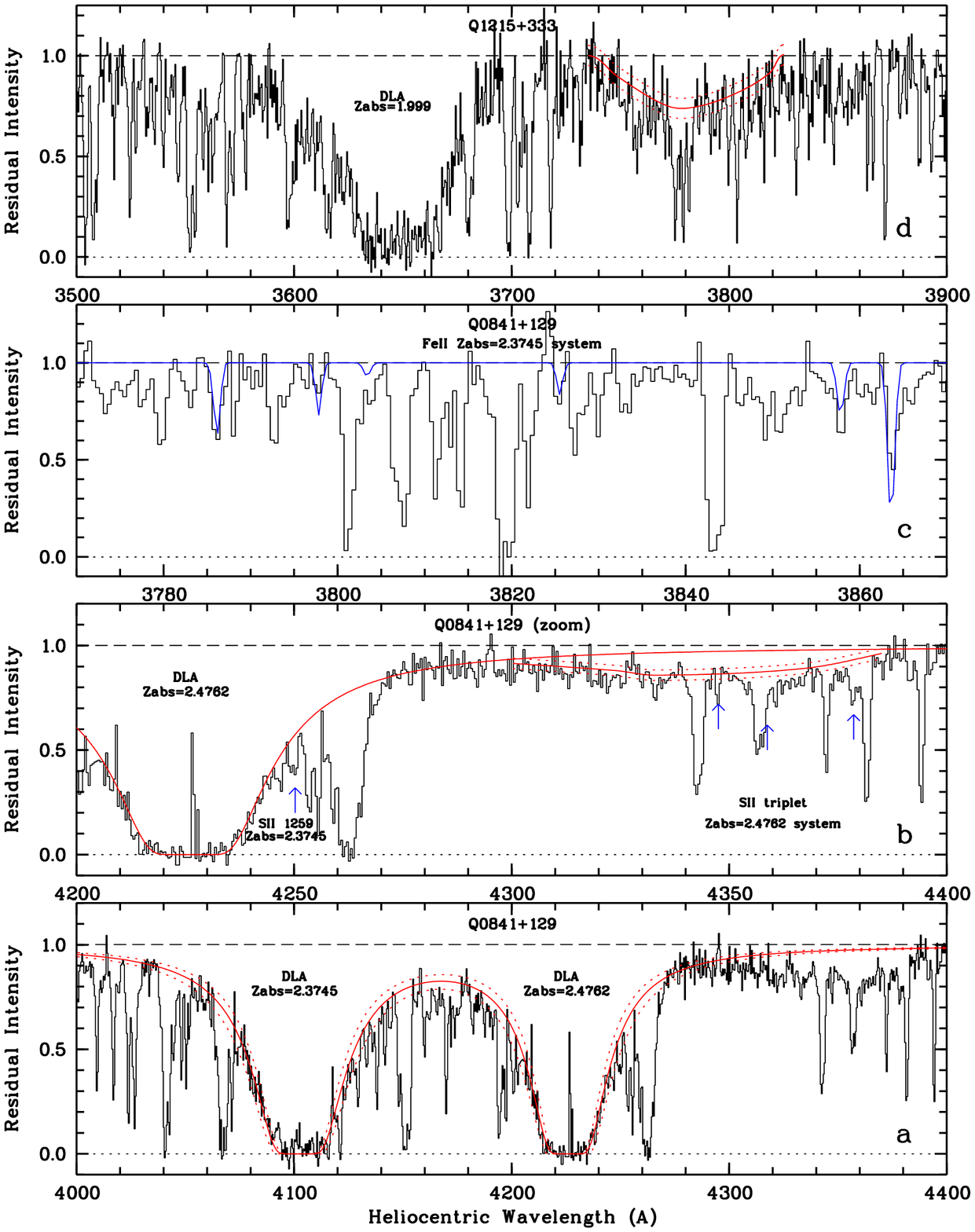,height=\vsize}  
\psfig{figure=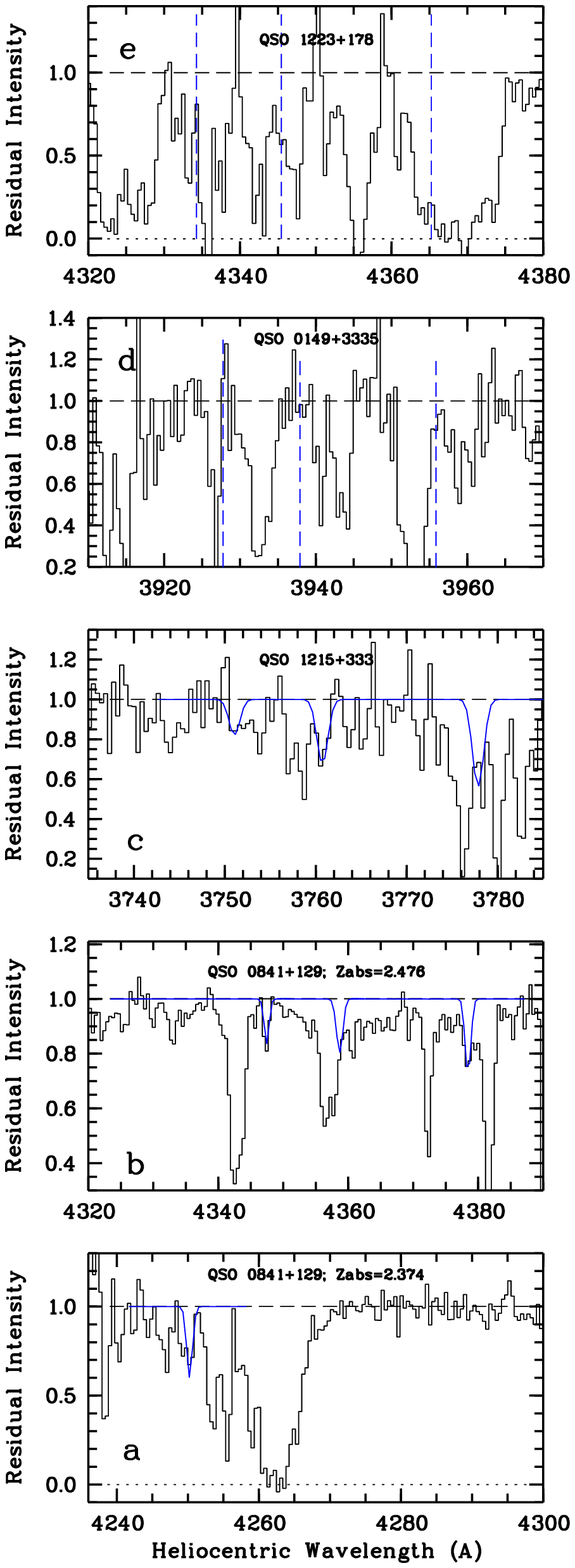,height=\vsize}  
\psfig{figure=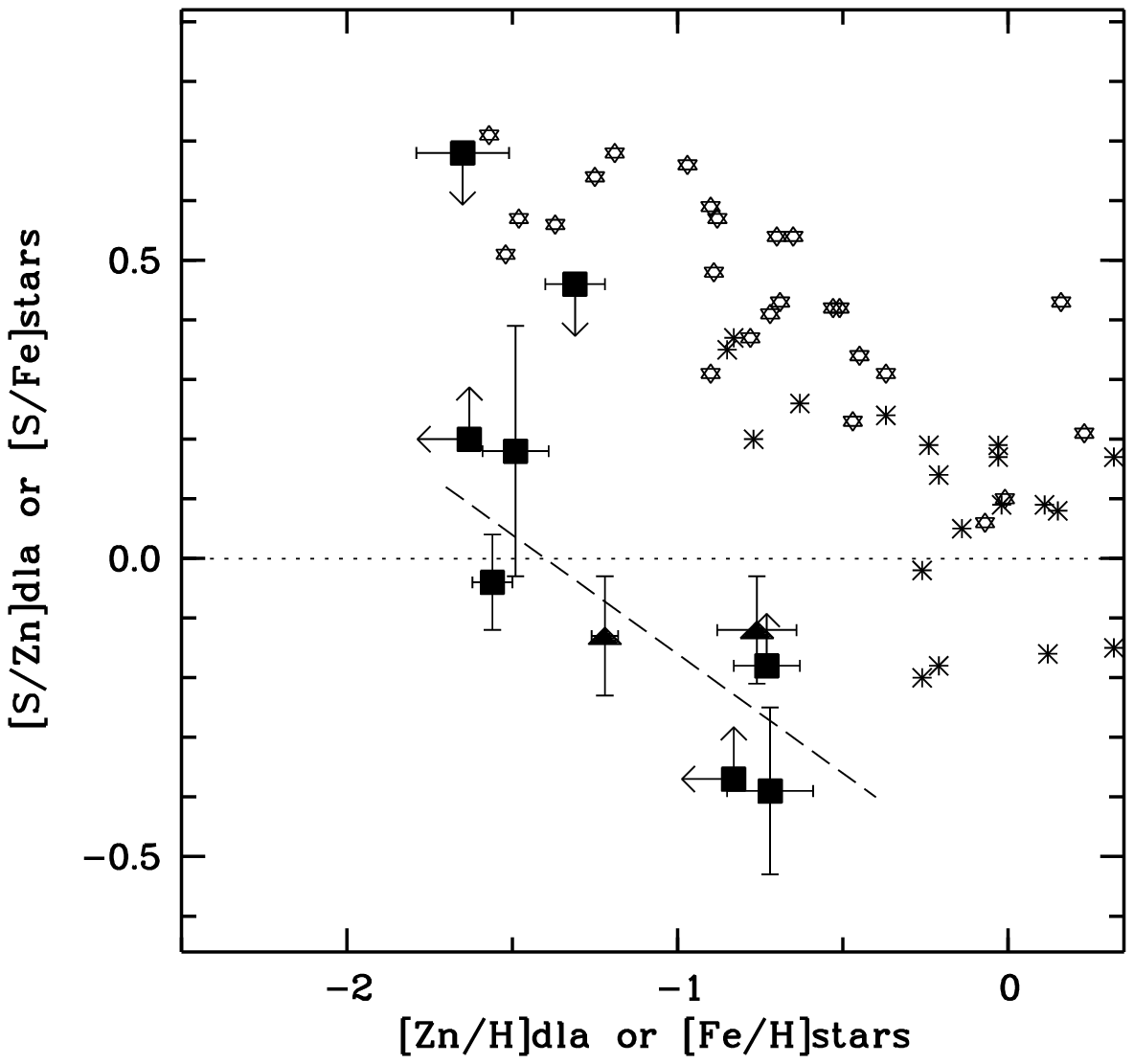} 
\psfig{figure=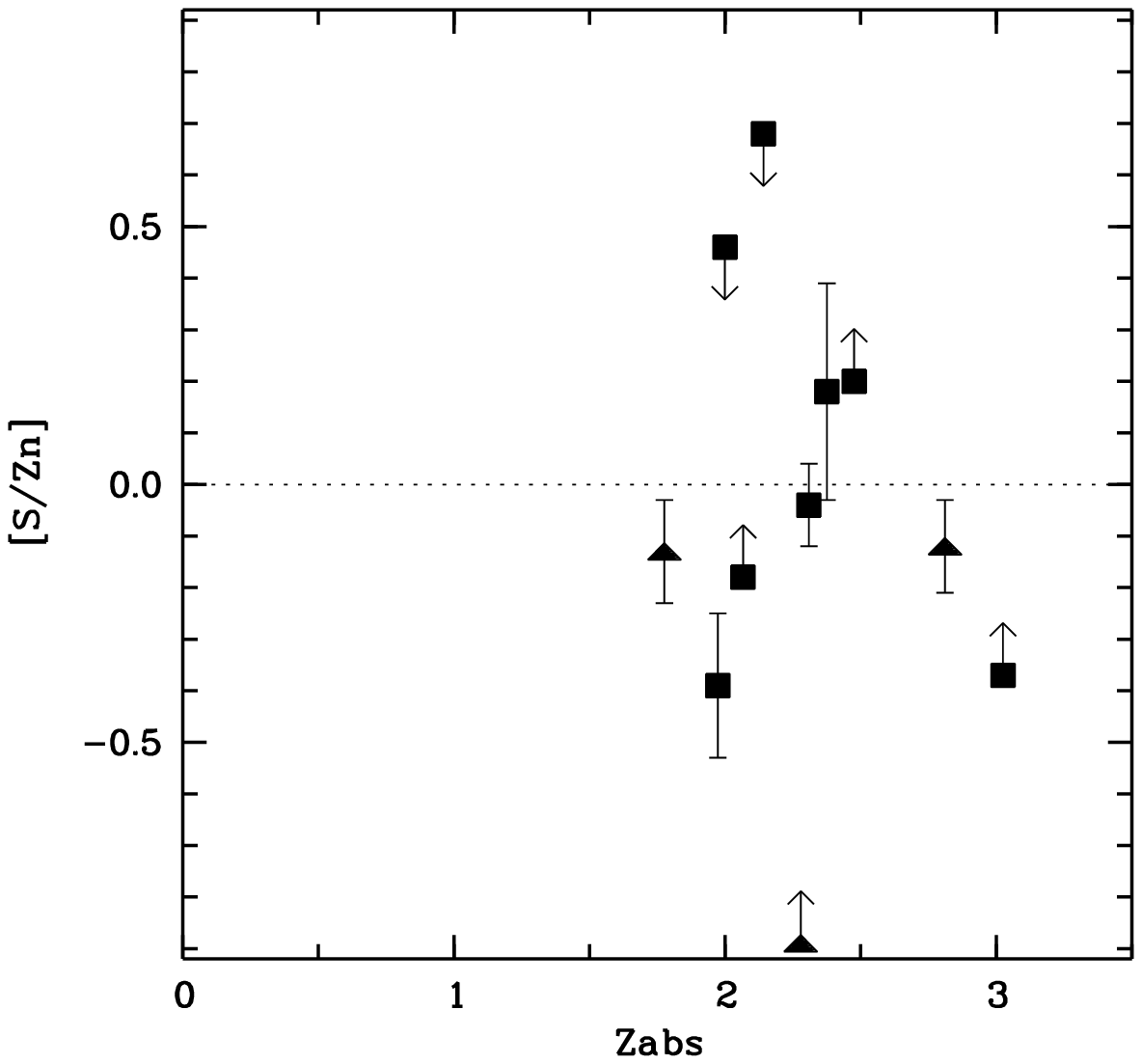}
  
\clearpage


\pagestyle{empty}
 
\begin{deluxetable}{lcccccccc} 
\footnotesize
\tablecaption{Journal of observations} 
\tablewidth{0pt}
\tablehead{
\colhead{QSO}     & \colhead{V}  &  \colhead{$z_{\rm em}$}& \colhead{$z_{\rm abs}$}&
\colhead{Telescope}&  \colhead{Date}    & 
\colhead{t$_{exp}$ (s)}&
\colhead {No. spectra} &
\colhead {Coverage (\AA)}}
\startdata
0013$-$004 & 18.2  & 2.084  & 1.9730 & ESO 3.6m &   1997 Sep 4  & 13500 & 2 & \nl
           &       &        &        &          &   1997 Sep 5  & 14400 & 2 & 3661-4638\nl
           &       &        &        &          &   1997 Sep 6  & 14400 & 2 & \nl
\nl	   
0149+335   & 18.5  & 2.431  & 2.1408 & WHT      &   1998 Dec 21 & 3600  & 2 & 3431-4366 \nl
\nl
0841+129   &  17.0 & 2.5:   & 2.3745 & WHT      &   1998 Dec 21 & 3600  & 2  & 3710-4643 \nl
           &       &        & 2.4762  & WHT     &   1998 Dec 23 & 4000  & 2  & 3710-4643 \nl
\nl 
1215+333   & 17.5  & 2.606  & 1.999  & WHT      &   1998 Dec 22 & 3560  & 2  & 3350-4190 \nl	   
	   &       &        &         &         &   1998 Dec 24 & 3600  & 2 \nl
\nl
1223+178   &  18.1 & 2.936  & 2.4658  & WHT     &   1998 Dec 24 & 3600  &  2 & 3880-4775 \nl
\nl   	   
2231$-$0015& 17.4  & 3.020  & 2.0662 & ESO 3.6m &   1997 Sep 5  &  7200 & 1  & 3661-4638 \nl
           &       &        &        &          &   1997 Sep 6  &  7200 & 1  &             \nl 
\enddata

\end{deluxetable}
\clearpage


\pagestyle{empty}
\begin{deluxetable}{lcccccccc}
\footnotesize
\tablewidth{-5cm} 
\tablecaption{Column densities.} 
\tablehead{
\colhead{QSO}   & \colhead{$z_{\rm abs}$}&
\colhead{log N(SII)}  &\colhead{b} & \colhead{ref} & 
\colhead{log N(ZnII) }  & \colhead{ref} &
\colhead{log N(HI)}  & \colhead{ref}
} 
\startdata
0013$-$004& 1.9730 &   14.86$\pm$0.12  & 24.5$\pm$10 & 1    &  
12.63$\pm$0.06 & 2 & 20.70$\pm$0.05 & 3   \nl
\nl
0149+3335 & 2.1408 &   $\leq$ 14.80    & ...         &  1   &  
11.50$\pm$0.10 & 4 & 20.50$\pm$0.10 & 4 \nl   
\nl
0841+129  & 2.3745 &   14.92$^{+0.16}_{-0.21}$     &  20$\pm$3\tablenotemark{a} &  1   &  
12.12$\pm$0.05 & 4  & 20.96$\pm$0.10 & 1 \nl
\nl
0841+129  & 2.4762 &   14.81$\pm$0.21  &  13.5$\pm$2 & 1    & 
$<$ 11.78      & 4    & 20.76 $\pm$0.10  & 1 \nl
\nl 
1215+333  & 1.999  &   $<$15.36  & 18-70      & 1 &    
12.29$\pm$0.06 & 4  & 20.95$\pm$0.07  & 2\nl  
\nl
2231$-$0015 & 2.0662  &  $>$ 14.90 &  \tablenotemark{b}& 1  &  
12.46$\pm$0.02 & 4 & 20.56$\pm$0.10 & 4               \nl
\enddata

\tablenotetext{a}{From the fit to the FeII transtions 1121,1125,1127,1133,1143,1144\AA}
\tablenotetext{b}{Limit obtained by using the linear part of the COG, 
since no information is available on b value, 
(see text for details)}
\tablenotetext{}{ REFERENCES ---
(1) This paper; (2) Pettini et al. (1994); (3) Ge \& Betchold (1997); (4) Prochaska \& Wolfe 1999 
}
\end{deluxetable}

\clearpage 


\pagestyle{empty}
\begin{deluxetable}{lccccccc}
\tablewidth{-4cm}  
\tablecaption{Abundances of $\alpha$ elements and iron-peak elements} 
\tablehead{
\colhead{QSO}    &  \colhead{$z_{\rm abs}$}&
\colhead{[S/H]\tablenotemark{a}}  & 
\colhead{[Si/H]\tablenotemark{b}} & 
\colhead{[Fe/H]} &
\colhead{[Cr/H]} &
\colhead{[Zn/H]} & 
\colhead{ Refs.\tablenotemark{c}}     
}  

\startdata
0013$-$004&  1.973   & --1.11$\pm$0.14            &  ...            & --1.84$\pm$0.05    & $\leq$--1.68      
& --0.72$\pm$0.13  & (1,2,2)  \nl
\nl  
0149+3335 & 2.141    & $\leq$--0.97               & --1.68$\pm$0.11  &  --1.81$\pm$0.10     & --1.37$\pm$0.11  
&--1.65$\pm$0.14   & (3,3,3)     \nl 
\nl             
0841+129  & 2.374    & --1.31$^{+0.19}_{-0.23}$   & --1.27$\pm$0.09  &  --1.64$\pm$0.17     & --1.56$\pm$0.10  
&--1.49$\pm$0.10   & (4,3,3)    \nl
\nl              
0841+129  & 2.476     & --1.22$\pm$0.23           &     $\geq$--1.85 &  --1.84$\pm$0.10     & --1.60$\pm$0.11                 
& $\leq$--1.63     & (3,3,3)     \nl
\nl              
1215+333  & 1.999     & $<$ --1.37                   & --1.47$\pm$0.08  &  $\geq$--1.81\tablenotemark{d}  &  --1.50$\pm$0.08         
&--1.31$\pm$0.09   & (3,3,3) \nl                                                                     
\nl
2231$-$0015&2.066  &$\geq$--0.93                       & --0.86$\pm$0.10  &  --1.32$\pm$0.10     & --1.07$\pm$0.11  
&--0.75$\pm$0.10   & (3,3,3) \nl
\enddata
 
\tablenotetext{}{NOTES---All abundances have been normalised to the meteorite
values reported by Anders \& Grevesse (1989): log (S/H)$_{\sun}$=--4.73$\pm$0.05,
log (Si/H)$_{\sun}$=--4.45$\pm$0.02,log (Fe/H)$_{\sun}$=--4.49$\pm$0.01, 
log (Cr/H)$_{\sun}$=--6.32$\pm$0.03,
log (Zn/H)$_{\sun}$=--7.35$\pm$0.02. Errors in [X/H]=log (X/H)$_{obs}$--log (X/H)$_{\sun}$ 
include errors in X and H column densities and errors in solar abundances.}
\tablenotetext{a}{From SII column densities obtained in this paper}
\tablenotetext{b}{From SiII column densities given in Prochaska \& Wolfe 1999}
\tablenotetext{c}{References for the column densities of FeII, CrII, \& ZnII respectively}
\tablenotetext{d}{Prochaska \& Wolfe 1999 gave log N(FeII)=14.65$\pm$0.04 derived 
from the saturated transition FeII 1608, and they warn that N(FeII) must 
be underestimated, hence we use this value as a lower limit}
\tablenotetext{}{REFERENCES---(1) Ge \& Betchold (1997); (2) Pettini et al. (1994); (3) Prochaska \& Wolfe 1999; (4) This paper}

\end{deluxetable}

 \clearpage 

\pagestyle{empty}
\begin{deluxetable}{lcccc}
\tablewidth{-4cm}  
\tablecaption{[$\alpha$/iron-peak] abundance ratios in DLAs } 
\tablehead
{
\colhead{QSO}    &  \colhead{$z_{\rm abs}$}&
\colhead{[S/Zn] }  & 
\colhead{[Si/Fe]} &
\colhead{[Si/Fe]$_{\rm corr}$}
}  
\startdata
0013$-$004& 1.973                     & --0.39 $\pm$0.14            &  ...		& ...
\nl
0100+1300 \tablenotemark{a} & 2.309  &  --0.04 $\pm$0.08            &  +0.37$\pm$0.07   & +0.15
\nl
0347-3819 \tablenotemark{b} & 3.025  & $\geq$ --0.37                & +0.31$\pm$0.03	& ...
\nl
0149+3335 & 2.141                     &  $\leq$ +0.68                & +0.13$\pm$0.04	& +0.03
\nl
0528--2505 \tablenotemark{c} & 2.811  &   --0.12$\pm$0.09            & +0.51$\pm$0.12	& +0.10
\nl
0841+129  & 2.374                     &   +0.18$^{+0.18}_{-0.21}$    & +0.37$\pm$0.24	& +0.23
\nl
0841+129  & 2.476                     & $>$ +0.20 \tablenotemark{d}& $>$ --0.01		&  ...
\nl
1215+333  & 1.999                     &  $<$ +0.46                 & $\leq$ +0.34$\pm$0.2& ...
\nl
1331+170 \tablenotemark{e} & 1.776    & --0.11$\pm$0.11             & +0.65$\pm$0.10    & +0.00
\nl
2231$-$0015&2.066                     & $\geq$ --0.18               & +0.46$\pm$0.03    & +0.04
\nl
2348$-$147 \tablenotemark{f} &2.279   & $\geq$ --0.89               & +0.37$\pm$0.02	&  ...
\nl
\enddata
\tablenotetext{}{NOTES--- Relative abundances [X/Y] have been directly obtained from
the column densities of X$^+$ and Y$^+$ species (references in Table 3, otherwise indicated)
and using the meteoritic values 
of Anders and Grevesse 1989. Errors in [X/Y] take into account error in X$^+$ and Y$^+$
column densities and errors in solar abundances.
[Si/Fe]$_{corr}$ are ratios corrected by dust effects following Vladilo (1998). 
}
\tablenotetext{a}{N(SII) from MCV98; N(ZnII), 
N(FeII) from Prochaska \& Wolfe 1999; N(SiII) from Lu et al 1998}
\tablenotetext{b}{N(SII) from Centuri\'on et al. 1998; N(ZnII) from Pettini et al. 1994; 
N(SiII) and N(FeII) from Prochaska \& Wolfe 1999}  
\tablenotetext{c}{N(SII), N(SiII), N(ZnII), N(FeII) from Lu et al. 1996}
\tablenotetext{d}{The most conservative value by using log N(SII)=14.81-0.21 see Table 2}
\tablenotetext{e}{N(SII) from Kulkarni et al 1996, N(ZnII), N(SiII), N(FeII) from 
Prochaska \& Wolfe 1999} 
\tablenotetext{f}{N(SII), N(SiII), N(FeII)  from Prochaka \& Wolfe 1999; N(ZnII) 
from Pettini et al. 1994;}			  
\end{deluxetable}

\end{document}